%% file: acl_latex.tex
\pdfoutput=1

\documentclass[11pt]{article}

\usepackage{booktabs,multirow,multicol,amsmath,amssymb}
\usepackage{graphicx}
\usepackage{url}
\usepackage{tikz}
\usepackage{tikz-network}
\usepackage{verbatim}
\usepackage{xcolor}
\usepackage{hyperref}
\usepackage{caption}
\usepackage{anyfontsize}
\usepackage{marvosym}

\newcommand\totalPapers{418K }
\newcommand\totalBroadLLMPapers{16,979 }
\newcommand\given[1][]{\:#1\vert\:}
\newcommand\enrichmentFigWidth{0.85}
\newcommand\suppEnrichmentFigWidth{0.9}

\usepackage{acl}

\usepackage{times}
\usepackage{latexsym}

\usepackage[T1]{fontenc}
\usepackage[utf8]{inputenc}

\usepackage{microtype}

\usepackage{inconsolata}

\title{Topics, Authors, and Institutions in Large Language Model Research: Trends from 17K arXiv Papers}

\newcommand{\aspace}{\hspace{.8em}}

\author{Rajiv Movva$^1$ \aspace
  Sidhika Balachandar$^1$ \aspace
  Kenny Peng$^1$ \\
   \bf{Gabriel Agostini$^1$ \aspace Nikhil Garg$^2$ \aspace Emma Pierson$^2$} \\ Cornell Tech \\
   \Letter: \texttt{rmovva@cs.cornell.edu} \\  
  \\}

\begin{document}
\maketitle

\def\thefootnote{1}
\footnotetext{Co-first authors.}
\def\thefootnote{2}
\footnotetext{Co-senior authors.}
\def\thefootnote{\arabic{footnote}}

\begin{abstract}

Large language models (LLMs) are dramatically influencing AI research, spurring discussions on what has changed so far and how to shape the field's future. 
To clarify such questions, we analyze a new dataset of 16,979 LLM-related arXiv papers, focusing on recent trends in 2023~vs.~2018-2022. 
First, we study disciplinary shifts: LLM research increasingly considers societal impacts, evidenced by 20$\times$ growth in LLM submissions to the Computers and Society sub-arXiv.
An influx of new authors -- half of all first authors in 2023 -- are entering from non-NLP fields of CS, driving disciplinary expansion.
Second, we study industry and academic publishing trends. 
Surprisingly, industry accounts for a \textit{smaller} publication share in 2023, largely due to reduced output from Google and other Big Tech companies; universities in Asia are publishing more. 
Third, we study institutional collaboration:~while industry-academic collaborations are common, they tend to focus on the same topics that industry focuses on rather than bridging differences.
The most prolific institutions are all US- or China-based, but there is very little cross-country collaboration.
We discuss implications around (1) how to support the influx of new authors, (2) how industry trends may affect academics, and (3) possible effects of (the lack of) collaboration.

\end{abstract}

\input{00_intro}

\input{02_methods}
\input{03a_results_topics}

\input{03b_results_authors}

\input{03c_results_industryacademia}
\input{03d_results_collaboration}
\input{01_relatedwork}
\input{04_discussion}

\bibliography{custom}

\appendix
\input{99_appendix}

\end{document}

%% file: 00_intro.tex
\section{Introduction}
\label{sec:intro}

Recent advances in language modeling have caused disruptive shifts throughout AI research, 
spurring conversation about how the field is changing and how it should change.
Discussions so far have drawn on surveys and interviews of NLP community members \citep{michael_what_2022, gururaja_build_2023, lee_surveying_2023}, and recurring themes under consideration include which topics are shifting in importance, which directions remain fruitful for academics and other compute-limited researchers, and which institutions hold power to shape LLM research.

In periods of flux like this one, \textit{bibliometrics} -- quantitative study of publication patterns -- offers a useful lens. 
Prior bibliometric analyses have been clarifying in NLP, identifying topic shifts \citep{hall_studying_2008}, flows of authors in and out of the field \citep{anderson_computational_2012a}, and the growing role of industry \citep{abdalla_elephant_2023}.
Due to the rapid recent growth of the LLM literature, fundamental questions about the topics, authors, and institutions driving its growth remain understudied.

Addressing this gap, we conduct a bibliometric analysis of recent trends in the LLM literature, focusing on changes in 2023 compared to 2018-22.
We collect and analyze 16,979 LLM-related papers posted to arXiv from Jan.~1, 2018 through Sep.~7, 2023. In addition to arXiv metadata, we annotate these papers with topic labels, author affiliations, and citation counts, and make all data and code publicly available.\footnote{\url{https://github.com/rmovva/LLM-publication-patterns-public}}
We analyze three questions: 

\begin{enumerate} 
\item \textbf{Which topics and authors are driving the growth of LLM research?} 
  Following prior work, we study topics and author movement as markers of a field's evolution \citep{kuhn_structure_1962, uban_studying_2021, anderson_computational_2012a}.
  In 2023, LLM research increasingly focuses on societal impacts: the Computers and Society sub-arXiv has grown faster than any other sub-arXiv in its proportion of LLM papers, up by a factor of 20$\times$ in 2023. 
  A more granular topic-level analysis echoes this result: the ``Applications of ChatGPT'' and ``Societal Implications of LLMs'' topics have grown 8$\times$ and 4$\times$ respectively. We also see rapid growth in other sub-arXivs outside of core NLP, including Human-Computer Interaction, Security, and Software Engineering.
  Simultaneously, BERT and task-specific architectures are shrinking due to centralization around newer models (e.g.,~GPT-4 and LLaMA).
    
    The increased focus on societal impacts and non-NLP disciplines is driven by a strikingly large proportion of authors new to LLMs. 
    Half (49.5\%) of LLM first authors in 2023 have never previously co-authored an NLP paper (and nearly two-thirds haven't previously co-authored an LLM paper), and a substantial fraction (38.6\%) of the last authors haven't either.
    These new authors are entering from other fields like Computer Vision, Software Engineering, and Security, and they are writing LLM papers on topics further out from NLP, e.g.,~vision-language models, applications in the natural sciences, and privacy/adversarial risks.
    
    \item \textbf{What are the roles of industry and academia?}
    The role of industry -- both what it is, and what it should be -- has been a topic of prior empirical measurement and normative discussion \citep{abdalla_elephant_2023, michael_what_2022}; the latest wave of LLM research has raised concerns of centralization around industry models and increased industry secrecy \citep{bommasani_foundation_2023}. 
    We identify two competing trends. 
    On one hand, industry publishes an outsize fraction of top-cited research, including widely-used foundation models and methodological work on topics like efficiency. 
    However, in 2023, industry is publishing \textit{less}: Big Tech companies are accounting for 13.0\% of papers in 2023, compared to 19.3\% before then, with a particular drop from Google.
    This trend suggests that reduced openness is not only playing out in high-profile cases (e.g.,~the opaque GPT-4 technical report \citep{rogers_closed_2023}), but is emerging as a broader, industry-wide phenomenon of reduced publishing.
    Academics, meanwhile, account for a larger share of papers (particularly universities in Asia). Relative to industry, more academic work applies models to non-NLP tasks and studies social impacts.

    \item \textbf{How are institutions collaborating?}
    Motivated by broader discussion around the risks of AI competition between nations and institutions~\cite{cuellar_ai_2023, hao_new_2023}, we analyze the network of collaborations between the 20 most prolific institutions, all of which are either US- or China-based.
    We document a US/China schism: pairs of institutions which frequently collaborate are almost exclusively based in the same country. 
    Microsoft, which collaborates with both American and Chinese universities, is the one exception.
    We also find that while industry-academic collaborations are common, rather than bridging differences, these papers skew significantly towards the topics usually pursued by industry.
    Collaborations between multiple companies are rare.
\end{enumerate}

How might these insights inform the NLP community, policymakers, and other stakeholders in the future of LLM research?
First, our analysis of topics and authors reveals that LLMs are increasingly being applied to diverse fields outside of core NLP.
Researchers performing interdisciplinary work should involve domain experts in both NLP and the other areas of study (e.g.~education, medicine, law); community leaders should reflect on how publication venues and peer review processes can best accommodate interdisciplinary work. 
We also show that LLM research is experiencing an influx of new authors, implying heightened value of educational resources, research checklists \cite{magnusson_reproducibility_2023}, and other frameworks to encourage good research practice \cite{dodge_show_2019, kapoor_reforms_2023}.
Second, we find that while industry continues to lead much of the most impactful research, large tech companies are publishing less overall. 
Academics lead important research on society-facing applications and harms of AI, but closed-source models hinder detailed evaluations \citep{rogers_closed_2023}.
Open-source datasets and models are therefore increasingly valuable, and the community should consider how to better incentivize these contributions. 
Third, we provide evidence of a lack of collaboration between the US and China, substantiating concerns about AI-related competition \cite{cuellar_ai_2023, hao_new_2023}. 
Institutions may have incentives that make collaboration difficult, but efforts to create consensus may help avoid unethical or risky uses of AI. 
By characterizing these recent changes in the LLM research landscape, our work aims to ground discussions on the policies and practices that will shape the field's future.

%% file: 02_methods.tex
\section{Methods}
\label{sec:methods}

We summarize our data and methods here and provide full details in Appendix \ref{sec:supplementary_methods}; Table \ref{tab:merged_rows} lists the fields we use in our analysis. 
Our primary dataset consists of all \totalPapers papers posted to the CS and Stat arXivs between January 1, 2018 and September 7, 2023. 
Following past ML survey papers \cite{fan_bibliometric_2023, peng_mitigating_2021, blodgett_language_2020, field_survey_2021}, we identify an analysis subset by searching for a list of keywords in paper titles or abstracts. Keyword search has the benefits of transparency, simplicity, and consistency with past work, but also has caveats; see \S \ref{sec:llm_related_definition} for further details. 
Our keyword list surfaces \totalBroadLLMPapers papers; the specific terms we include are \{\texttt{language model, foundation model, BERT, XLNet, GPT-2, GPT-3, GPT-4, GPT-Neo, GPT-J, ChatGPT, PaLM, LLaMA}\}. Details about this list are in  \S \ref{sec:llm_related_definition}.

We define several fields for each paper in this subset. In doing so, we follow past work and conduct manual audits to assess the reliability of our annotations; however, there remain inherent limitations in how these fields are defined, as we discuss in Appendix \ref{sec:supplementary_methods}.
We tracked a paper's primary sub-arXiv category, e.g.,~Computation and Language (cs.CL).
For more fine-grained topic analysis, we assigned each paper one of 40 LLM-related topics (\S \ref{sec:methods_topic_model}). We clustered embeddings of paper abstracts~\cite{zhang_neural_2022, grootendorst_bertopic_2022}, then titled the clusters using a combination of LLM annotation and manual annotation.
We annotated papers for whether their authors list academic or industry affiliations (\S \ref{sec:methods_industry_academic}). 
We pulled citation counts from Semantic Scholar \cite{kinney_semantic_2023}, and tracked the \emph{citation percentile} for each paper: the percentile of its citation count relative to papers from the same 3-month window (\S \ref{sec:methods_citation_counts}). 

%% file: 03a_results_topics.tex
\section{Results}

Past work has shown that the raw count of LLM papers has risen steeply~\cite{fan_bibliometric_2023, zhao_survey_2023}.
These trends replicate on arXiv: 12\% of all CS/Stat papers were LLM-related in mid-2023 (Figure \ref{fig:supp_percent_llm_arxiv}). 
Compared to all other topics (Figure \ref{fig:supp_top_arxiv_cs_topics}) and all words/bigrams in paper abstracts (Table \ref{tab:keyword_enrichment}), LLM- \& generative AI-related topics and terms are growing fastest.
We dissect these ongoing changes by studying the topics, authors, and institutions that are accounting for them.

\subsection{Which topics and authors are driving the growth of LLM research?}

\subsubsection{How have topics shifted in 2023?}
\label{sec:results_part_b_what_topics}

We begin by analyzing the changing topic distribution of language modeling research -- taxonomized by sub-arXiv category and semantic clusters -- to identify which threads within LLM research are expanding and shrinking fastest.

\begin{figure}[!ht]
    \begin{center}  \includegraphics[width=0.47\textwidth]{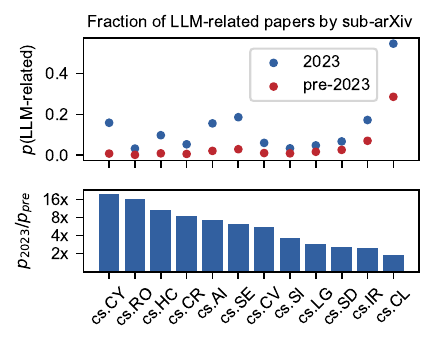}
    \end{center}
    \caption{\textbf{Sub-arXivs growing fastest in fraction of LLM papers.} 
    \textit{Top:} Proportions of LLM-related papers in a sub-arXiv in 2023 (blue) and pre-2023 (red).
    \textit{Bottom:} sub-arXivs are sorted by the ratio of these two quantities, representing how much more likely, in 2023, a random paper in the sub-arXiv involves language models.
    The $x$-axis labels correspond to, respectively: Computers and Society, Robotics, Human-Computer Interaction, Cryptography and Security, Artificial Intelligence, Software Engineering, Computer Vision, Social and Info.~Networks, Machine Learning, Sound, Information Retrieval, and Computation and Language.
    }
    \label{fig:subarxiv-growth}
\end{figure}

\paragraph{LLM papers increasingly involve societal impacts and fields beyond NLP.}
For a coarse analysis of where LLMs are growing fastest, we use a paper's designated primary sub-arXiv category.
We rank sub-arXivs by how quickly their proportion of LLM papers is increasing, i.e.,~according to the ratio $\frac{p(\text{LLM paper} \mid \text{paper on sub-arXiv \& 2023})}{p(\text{LLM paper} \mid \text{paper on sub-arXiv \& pre-2023})}$.
Figure \ref{fig:subarxiv-growth} displays all sub-arXivs (with at least 50 LLM papers) sorted by their 2023 to pre-2023 ratios.
Computers and Society (cs.CY) ranks first, with a ratio of 20$\times$: in 2023, 16\% of its papers are about LLMs, compared to just 0.8\% pre-2023.  
This society-facing work ranges widely, including discussions of the impacts of LLMs on education (\citealt{kasneci_chatgpt_2023, chan_students_2023}, inter alia), ethics and safety \citep{ferrara_should_2023, sison_chatgpt_2023}, and law \cite{henderson_foundation_2023, li_dark_2023}.
Other sub-arXivs with both rapid growth and at least 10\% prevalence of LLM papers in 2023 include HCI (up to 10\% of all papers in 2023), AI (16\%), and Software Engineering (19\%).
Strikingly, 55\% of Computation and Language (cs.CL) papers are LLM-related, but due to its already-large fraction before 2023 (29\%), its rate of increase ranks last.
LLMs are clearly impacting much of CS research beyond NLP, especially in society- and human-facing fields.

\paragraph{The fastest-growing LLM topics cover applications, capabilities, and methods.}
To study topic shifts at a more granular level, we observe the changing distribution of the 40 LLM-related topics with which we annotated the corpus. 
Since these topics are learned only on the LLM paper distribution, they are more specific than sub-arXivs, which span all of Computer Science.
Figure \ref{fig:fast-growing-topics} lists the five fastest growing and shrinking topics in 2023 according to $\frac{p(\text{topic} \given \text{published in 2023})}{p(\text{topic} \given \text{published pre-2023})}$, and the results corroborate the sub-arXiv analysis (full results in Table \ref{appendix:supp-enriched-topics-2023}). 
The fastest-growing topic is ``Applications of LLMs/ChatGPT'', which has risen from 0.9\% of LLM papers before 2023 to 7\% in 2023, an 8$\times$ increase.
This cluster of papers, along with ``Societal Implications of LLMs'' (4$\times$ growth, 4th fastest-growing), captures papers on empirical studies of LLMs for applied tasks, discussions of societal applications of ChatGPT, and ethical arguments (\citealt{dai_uncovering_2023, mogavi_exploring_2023, derczynski_assessing_2023}, inter alia).
The next two fastest-growing topics --- ``Software, Planning, Robotics'' and ``Human Feedback \& Interaction'' --- hint further at applications.
The former topic includes papers on two promising use cases of recent models, code generation and robotics, while the latter concerns the growing role of human feedback and HCI in developing useful language systems.
\paragraph{Shrinking topics highlight centralization around closed-source models.}
The ``BERT \& Embeddings'' topic is shrinking, consistent with prompt-based, few-shot models now replacing fine-tuned BERT systems.
Many papers in this shrinking topic also study internal model representations, which may be less common now that token probabilities and activations are inaccessible from widely-used closed-source models like GPT-4.
Topics on transformers, transfer learning, and language correction are also shrinking; recent, general-purpose models may have rendered some of this architecture-/task-specific research less relevant.
While centralization is not new to NLP -- nearly half of papers in recent years cited BERT \citep{gururaja_build_2023} -- the centralization around \textit{closed-source} models specifically may harm scientific practice \citep{rogers_closed_2023}.

%% file: 03b_results_authors.tex
\subsubsection{Who are the authors driving the expansion of LLM research?} 

Based on our findings that LLMs research is expanding outside of core NLP, it is natural to ask \textit{who} is driving the broadening of the field. 

\paragraph{In 2023, nearly half of LLM first authors have not previously published on NLP.}
To what extent are researchers from non-NLP fields responsible for the growth of LLM research?
For each LLM paper in our corpus, we ask whether the first author of the paper had previously written an NLP paper on arXiv -- that is, any co-authored paper for which cs.CL was listed as one of the paper's sub-arXiv categories (\S\ref{sec:methods_author_background}). 
We similarly code papers according to their last authors. 
We refer to LLM authors without a prior NLP paper as \textit{new}, and authors with a prior NLP paper as \textit{experienced}.

Consistent with the broader growth of the field, the raw counts of new first and last authors has increased consistently since 2018, with an especially large jump in 2023 (Figure \ref{fig:lineplot-new-authors}). 
More surprisingly, however, the \textit{percentage} of authors without NLP background has also increased this year: in 2022, the percentages of new first and last authors were 41\% and 29\% respectively, and are up to 50\% and 39\% in 2023.
Figure \ref{fig:lineplot-new-authors} plots these statistics each year since 2018. 
The last time the percentage of new authors was this high was in 2018, which was another inflection point as language models started to become useful for many tasks (e.g.~ELMo, BERT).
In the four years following 2018, the field grew significantly, but the percentage of authors without an NLP background was constant or declining. 
The increase in new author percentage in 2023, then, reflects that more LLM authors are either new to research entirely or are moving into LLM research from other fields.
Seeing as both first authors and last authors are more likely to be new, the field is widening in its composition both of junior researchers and senior researchers.
Next, we analyze the contributions of these first-time authors further, to understand both their disciplinary origins and the types of contributions they are making to LLM research.


\paragraph{What fields are new authors coming from?} 
Out of the 2,746 unique first authors in 2023 who had no prior NLP paper, 57\% of them had at least one paper in another field.
We compare the publication histories of these new authors to experienced authors (Table \ref{tab:past-subarxivs}; further details in Appendix \ref{sec:pub_history}).
About half of new authors' prior papers are in Computer Vision and Machine Learning, which are common categories overall.
But a long tail of less common sub-arXivs further distinguishes new and experienced authors: new authors have published more in Software Engineering, Robotics, Security, and Social Networks, which are the same sub-arXivs growing in overall LLM proportion in 2023 (\S\ref{sec:results_part_b_what_topics}).



\paragraph{New authors are driving the increased disciplinary diversity of LLM research.} 
We close the loop by studying the LLM-related papers being written in 2023 by new authors.
Confirming intuition, LLM papers written by authors without a prior NLP paper are more likely to fall in non-NLP sub-arXivs, including Software, Computers and Society, HCI, and Security (Table \ref{supp:tab_2023_newauth_llm_subarxiv}).
Comparing paper topic distributions, new authors are more likely to publish on ``Visual Foundation Models,'' ``Applications of LLMs/ChatGPT,'' and ``Natural Sciences,'' while experienced authors publish more on ``Interpretability \& Reasoning,'' ``Knowledge Distillation,'' and ``Summarization \& Evaluation'' (additional topics in Figure \ref{appendix:topics_experience_first_author}). 
That is, authors without NLP background are using LLMs in their home fields, while authors with past NLP background continue to focus more on improving, understanding, and evaluating LLMs.

Overall, there is a clear trend that authors without prior NLP experience are accounting for a substantial and increasing fraction of LLM-related research contributions. 
These new authors include both junior and senior researchers, many of whom have previously published in other fields, and are widening the scope of LLM-related research directions.
The influx of new authors is clearly valuable, but also creates room for error or miscommunication: the large number of junior researchers entering LLM research may not be familiar with prior work, and the broader use of LLMs outside NLP may result in improper usage or experimental practice \cite{narayanan_evaluating_2023}. 
Onboarding resources, research checklists, and other standards can help maintain good practice despite growth \cite{magnusson_reproducibility_2023, kapoor_reforms_2023}.

%% file: 03c_results_industryacademia.tex
\subsection{What are the roles of industry \& academia?}

We turn our focus from authors to larger-scale institutional trends. 
To study institutions, we annotated each paper with author affiliations extracted from full-texts (\S\ref{sec:methods_industry_academic}).
We manually coded every institution with at least 10 papers as either academic or industry,\footnote{Eight institutions did not fall clearly into this binary, \textit{e.g.}, nonprofits and government labs. AllenAI was the only such institution in the top 100 producers.} resulting in 280 academic and 41 industry institutions.
Out of the 14,179 papers with at least one extracted affiliation, 11,627 (82.0\%) were written by one of these 321 institutions.

\paragraph{A large (and growing) majority of LLM papers are published by academic institutions.} 

Figure \ref{fig:top_institutions} displays the institutions with the most LLM-related papers.
Several Big Tech companies lead in paper count: Microsoft and Google are top two, and Amazon, Meta, and Alibaba are also in the top 10.
However, overall, industry institutions account for a relatively small fraction of research output: just 32\% of papers have an industry affiliation (and over half of these are industry-academic collaborations).
85\% of papers have an academic affiliation, and 41 out of the top 50 institutions are academic, led by CMU, Stanford, Tsinghua, Peking, and UW.
Further, academic LLM research has grown \textit{faster} in 2023 than has industry research: there are 3.3$\times$ as many academic as industry papers in 2023, compared to 2.3$\times$ pre-2023.
This result is somewhat surprising in light of the longer-term trend of greater industry presence in NLP and AI \citep{abdalla_elephant_2023, birhane_values_2022}, suggesting that the shift we observe is specific to LLM research in 2023.

\begin{table}
\small
\centering

\caption{\textbf{Institutions with the largest changes in LLM publishing fraction in 2023.} 
The five institutions with the largest increases and largest decreases in publication share in 2023 vs. pre-2023 (2018-22). 
All changes were significant at $p < 0.05$ under a $\chi^2$ test, after multiplying $p$-values by $10$ to account for multiple testing.
}

\label{tab:changing_institutions}

\begin{tabular}{lccc}
\toprule
            Institution & pre-2023 & 2023 & Diff. \\
\midrule
               Zhejiang &     0.9\% & 2.1\% &   +1.2\% \\
          NTU Singapore &     0.8\% & 1.9\% &   +1.1\% \\
Chinese U. of Hong Kong &     0.6\% & 1.7\% &   +1.1\% \\
Chinese Academy of Sci. &     1.7\% & 2.7\% &   +1.0\% \\
                    NUS &     0.8\% & 1.6\% &   +0.8\% \\
\midrule
                 Google &     6.7\% & 3.8\% &  -2.9\% \\
              Microsoft &     6.8\% & 5.4\% &  -1.4\% \\
                 Amazon &     3.0\% & 1.9\% &  -1.1\% \\
               Meta &     2.9\% & 1.9\% &  -1.0\% \\
                     UW &     2.6\% & 1.7\% &  -0.9\% \\
\bottomrule
\end{tabular}

\end{table}

\paragraph{Big Tech companies are publishing less, and universities in Asia are publishing more.}
Examining which specific institutions are publishing more and less reveals two specific trends on the changes in publication share (Table \ref{tab:changing_institutions}).
Strikingly, out of all 321 institutions, the top four Big Tech publishers -- Google, Microsoft, Amazon, and Meta -- are the four institutions with the largest decreases in publication fraction in 2023. 
Collectively, they accounted for 19.3\% of all LLM papers pre-2023, down to 13.0\% in 2023.
Google has shown a particularly large drop: its share of LLM papers roughly halved in 2023 relative to prior years. 
Reduced industry publishing in 2023, especially from Big Tech companies, may have multiple explanations, such as a deprioritization of basic research or heightened secrecy due to competition. 
Companies releasing fewer modeling details has been previously noted, but only in high-profile cases like the GPT-4 technical report \cite{rogers_closed_2023}.\footnote{And recently, Gemini: \small{\url{https://x.com/JesseDodge/status/1732444597593203111}}.}
Our data suggest that reduced openness is emerging as a broader phenomenon, where less is being published \textit{at all} by top companies.

All of the top ten institutions with the largest increases in publication share are academic institutions, concordant with the overall rise in academic percentage. 
Notably, they are all institutions in Asia (China, Singapore, Hong Kong, UAE). 
These ten institutions account for 16.4\% of papers in 2023, up from 9.1\% pre-2023.
In contrast, the top ten US universities have not changed significantly in their publishing fraction (18.8\% to 18.0\%). 
The growth of LLM research from universities in Asia deserves further consideration, especially given prior discussions of citation gaps between countries \citep{rungta_geographic_2022} and the geopolitics around LLM development \citep{ding_recent_2023}. 

\begin{figure}
    \begin{center}  \includegraphics[width=0.47\textwidth]{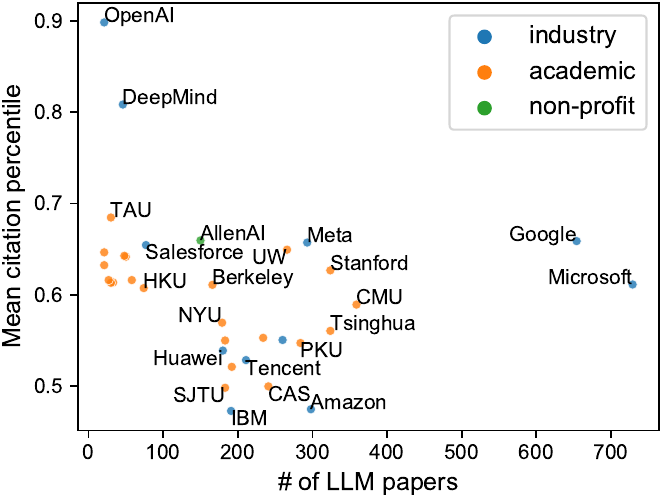}
    \end{center}
    \caption{Mean citation percentile vs. number of LLM papers for the 34 institutions which are in the top 20 of either metric. 
    Some point labels were removed for visual clarity. 
    Citation percentile is defined in \S\ref{sec:methods}.
    }
    \label{fig:citations-vs-count}
\end{figure}

\paragraph{OpenAI and DeepMind publish infrequently with high-impact.} 
Paper count is an imperfect metric of influence: there are outlier institutions that rarely publish, but have large impact when they do.
In Figure \ref{fig:citations-vs-count}, we plot mean citation percentile against overall paper count to identify some of these institutions.
OpenAI and DeepMind are the largest outliers: they both publish infrequently, but receive many citations. 
OpenAI has just 21 papers as annotated in our corpus, but, on average, they are in the 90th percentile of citation counts.
Other institutions like Google, Meta, UW, and AllenAI both publish frequently and receive high citation counts.

\begin{figure*}[!htb]
    \begin{center}  \includegraphics[width=\enrichmentFigWidth\textwidth]{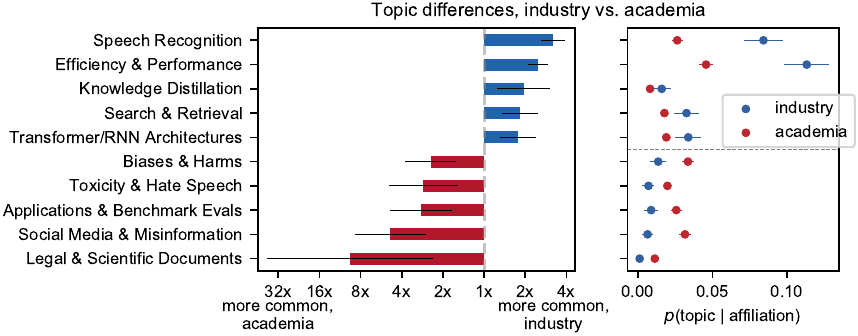}
    \end{center}
    \caption{Topics which occur most disproportionately among industry (blue) vs.~academic (red) papers.
    \textit{Left:} Topics are sorted by the ratio $p(\text{topic} \mid \text{industry-only}) / p(\text{topic} \mid \text{academic-only})$, excluding industry-academic collaboration papers and papers with no inferred affiliations. \textit{Right:} Topic frequencies by group.
    }
    \label{fig:topics-industry}
\end{figure*}

\paragraph{Top-cited papers are split between industry and academia.}
We manually annotated the top 50 cited papers in 2023 (as of September): 30 papers have an industry affiliation, 30 papers have an academic affiliation, and 10 have both.
The fact that 60\% of top-cited papers have an industry affiliation -- compared to 27\% of papers overall in 2023 -- shows that industry research continues to strongly influence the field, even if companies are publishing less overall.
We also annotated these 50 papers for whether they are model-focused, \textit{i.e.}~reporting on a newly trained language/foundation model, or evaluation-focused, \textit{i.e.}~evaluating a model on new tasks or applications (for example, \citealt{nori_capabilities_2023, guo_how_2023, gilardi_chatgpt_2023}). 
23 papers were model-focused, of which the majority (19) had an industry affiliation.
17 were evaluation-focused, of which the majority (12) had an academic affiliation.
These two categories of top-cited papers allude to the relative specializations of industry and academia, which we explore further.

\paragraph{Industry focuses more on general-purpose methods, while academics apply these models.} 
To understand their publishing differences at scale, we compute the topics most skewed towards either industry or academia.
Figure \ref{fig:topics-industry} illustrates the five topics with the largest ratios of $\frac{p(\text{topic} \given \text{industry})}{p(\text{topic} \given \text{academic})}$ in blue, and the five with largest $\frac{p(\text{topic} \given \text{academic})}{p(\text{topic} \given \text{industry})}$ in red.\footnote{Papers resulting from collaboration are excluded from this analysis and studied later.} 
Industry papers are $\sim$2-3$\times$ more likely to cover methodological contributions, especially involving efficiency or model architecture. 
They also work more on certain tasks (speech recognition, search, retrieval) which may be more commercially relevant, and therefore more relevant to industry practitioners.
Academic papers more often apply LLMs to downstream, society-facing tasks, like legal document analysis, social media data, and hate speech detection.
They also prioritize studying biases \& harms more than industry.
Finally, academics are 3$\times$ as likely to write on ``Applications \& Benchmark Evals.'' 
These topic differences support the idea that industry is leading the way on model development research, while academics are using models for applied tasks and studying their implications.

%% file: 03d_results_collaboration.tex
\subsection{What are the patterns of collaboration?}
\label{sec:results_part_f_international_collaboration}

We examine which institutions tend to co-author papers together, focusing in particular on patterns of academic/industry and international collaboration. 

\paragraph{Academic-academic and industry-academic collaborations are both frequent, while industry-industry collaborations are rare.}
We consider collaborations among those institutions with at least 10 LLM papers. There are 2,250 papers produced by a collaboration between at least two academic institutions, and 2,084 by at least one academic institution and one industry institution; 55\% of all industry papers involve an academic collaboration. There are only 104 papers by a collaboration between at least two industry institutions. 

\paragraph{Industry-academic collaborations typically align with industry topics.}
We compute the topics that are most common in industry-academic collaborations relative to the baseline topic frequencies.
Similarly, we compute baseline-adjusted topic frequencies for industry-only author teams and academic-only author teams. 
We find that collaborations focus on very similar topics as industry-only teams: their normalized topic distributions have Spearman $\rho = 0.66, p < 10^{-5}$.
Meanwhile, collaboration topics are as different from academic topics ($\rho = -0.87$) as industry topics are from academic topics ($\rho = -0.92$).
For example, both industry-only and industry-academic papers focus on efficiency, translation, and vision-language models, while both groups publish infrequently on social media \& misinformation, bias \& harms, and interpretability.
These results are natural given that many collaborations result from PhD interns or student researchers, where company priorities may hold more weight. 
However, as compute and modeling resources continue to separate academics from researchers at Big Tech companies \citep{lee_surveying_2023}, stakeholders should consider where collaborations -- most of which involve Big Tech companies -- can be recast as an opportunity to bridge different expertise for mutual benefit. 

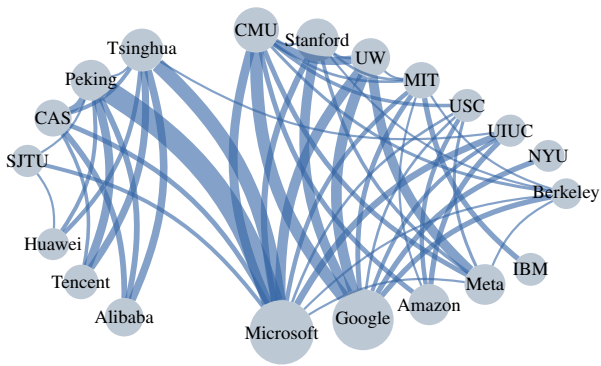
\begin{figure}[!ht]
\begin{center}
\resizebox{0.51\textwidth}{!}{

    \begin{tikzpicture}
    \Text[x=6.3, y=11.8, style={font=\Huge}]{Academic (U.S.)}
    \Vertices[IdAsLabel, RGB, style={opacity=0.3, draw=none}]{figures/vertices.csv}
    \Edges[RGB, opacity=0.6]{figures/edges.csv}
    \Text[x=-5.6, y=11.8, style={font=\Huge}]{Academic (China)}
    \Text[x=6.3, y=1.96, style={font=\Huge}]{Industry (U.S.)}
    \Text[x=-5.6, y=1.96, style={font=\Huge}]{Industry (China)}
    \end{tikzpicture}
    }
    \caption{Collaborations between the 20 institutions with the most LLM papers. 
    Node area is proportional to number of papers and edge width to number of collaborations between nodes (we show only edges corresponding to $\ge 5$ collaborations). 
    Microsoft collaborates with academic institutions across the U.S. and China (UIUC also has exactly 5 papers with Tsinghua).
    There are several notable academic collaborations and industry-academic collaborations, especially involving Microsoft, Google, UW, CMU, Stanford, Tsinghua, and Peking.
    }
    \label{fig:colab-network}
\end{center}
\end{figure}

\paragraph{Collaborations between American and Chinese institutions are rare.}
In Figure \ref{fig:colab-network}, we present the network of the 20 institutions that produced the most LLM papers in our dataset, all of which are China- or US-based. 
An edge between two institutions indicates that at least five papers resulted from a collaboration between them.
The network suggests little collaboration between the US and China with the exception of Microsoft, which collaborates with institutions in both countries.
Excluding Microsoft, the average number of collaborations between top-20 US and Chinese institutions is 1.1 papers, while the average edge weight between two US institutions is 6.2 papers and between two Chinese institutions is 7.6.
There is also only one edge between a pair of companies (Microsoft-Meta), supporting the prior statistic that industry-industry collaborations are relatively rare.

%% file: 01_relatedwork.tex
\section{Related Work}

Our work applies methods from bibliometrics and science of science \cite{borgman_scholarly_2002, fortunato_science_2018} to understand the emerging LLM literature.
These methods have a rich history in NLP; since the 2008 release of the ACL Anthology, several papers have applied quantitative methods to trace topic changes and paradigm shifts in the Anthology over time \cite{hall_studying_2008, anderson_computational_2012a, mohammad_state_2019, pramanick_diachronic_2023}.  
Another line of work focuses specifically on citation patterns: \citet{wahle_we_2023} study cross-disciplinary cites from NLP to other fields, \citet{singh_forgotten_2023} show that NLP papers are citing increasingly recent work (and less older work), and \citet{rungta_geographic_2022} identify geographic disparities in citation counts of published papers. 

Some bibliometrics papers also ask similar research questions to ours. 
\citet{abdalla_elephant_2023} study Big Tech companies in NLP research, identifying key players and showing significant growth in overall industry presence since 2017.
Their findings complement our more specific analysis of the language modeling literature, in which there are different trends than NLP overall (for example, we find that industry research actually \textit{declined} relative to academia in 2023).
\citet{fan_bibliometric_2023} apply bibliometric methods to a similar paper corpus, pulling 5,752 published language modeling papers from Web of Science from Jan 2017 - Feb 2023. 
They perform topic modeling and analyze co-citation networks to taxonomize the LLM literature, and they also study cross-country collaboration.
Our work differs in that we focus on specific changes in 2023 compared to years prior (rather than analyzing the entire time period collectively), and we use the arXiv in order to capture a recent sample with less publication delay.
Our empirical analyses also cover different questions, with more focus on emerging topics, first-time authors, and industry/academia dynamics.

Other related work uses qualitative methods to understand recent shifts in NLP. \citet{gururaja_build_2023} present a retrospective analysis centered on 26 interviews with NLP researchers, identifying key trends like the rise of benchmarks, software centralization, and reduced barriers to enter NLP. 
\citet{michael_what_2022} report on the results of the NLP community meta-survey, in which 327 researchers were polled for their views on the status and future of NLP.
\citet{saphra_first_2023a} draw historical analogy from the convergence around LLMs to the early statistical machine translation era, suggesting lessons and evergreen research directions.
It is encouraging that many other researchers have approached similar questions; together, the rich variety of datasets and methods can help us parse the ongoing expansion and flux of NLP.

%% file: 04_discussion.tex
\section{Discussion}
\label{sec:discussion}

We conduct a bibliometric analysis documenting recent changes and trends in the LLM research landscape. Our findings invite discussion on how researchers and policymakers should react to these changes. Drawing on our findings, we close with open questions for community reflection:
\begin{itemize}
    \item We show that LLM research increasingly involves disciplines beyond core NLP. How can we foster effective interdisciplinary research?
    \item We show there is a large influx of authors new to LLM research. How can we efficiently onboard these new authors and encourage high-quality research practice?
    \item We show that centralization around industry models is increasing even as companies are publishing less openly, reducing transparency. How can we incentivize open-source contributions or empower academics to be less reliant on industry?
    \item We show schisms in the collaboration network between the US and China, and between companies. How do we avoid harmful effects of competition and/or encourage cross-country \& cross-company collaboration?
\end{itemize}

\section*{Limitations}

Our analyses have limitations.
First, much of our work focuses on changes in the first nine months of 2023, a relatively short time period. 
In future work, we can re-apply our (publicly available) analysis toolkit to updated samples of the literature, or to adjacent fields of AI research.
Second, arXiv does not capture all LLM publications. 
We chose arXiv over the ACL Anthology to include fields beyond NLP and unpublished work (some companies post to arXiv but do not publish, for example), and because most ACL Anthology papers are posted to arXiv,\footnote{Out of a random sample of 50 LLM papers in the ACL Anthology, only 6 (12\%) were not on arXiv.} but the reverse is not true.
However, future work should study other scholarly corpora, like Semantic Scholar, which may cast a wider net \cite{kinney_semantic_2023}.
Third, our analyses rely on imperfect labels -- \textit{e.g.}, the topic of a paper, whether it is LLM-related, and whether authors have an academic/industry affiliation. 
All these variables are, for reasons we document in \S \ref{sec:methods} and Appendix \ref{sec:supplementary_methods}, potentially observed with both bias and noise, though we perform manual analyses where possible to understand sources of error.
Fourth, while we find very little US-China collaboration in LLM research, it is unclear how much baseline US-China collaboration there is across computer science in general. 
A recent analysis argues that the US and China have extensive historic collaborations in technical fields \cite{hao_turning_2023}, but future work should study US-China collaboration specifically in CS research to provide more context around our result.
However, there is reason to believe that irrespective of their previous collaboration or lack of collaboration in CS, it is especially important to encourage collaboration in LLM research, where competition and secrecy may hurt the entire field and harm society \cite{rogers_closed_2023, cuellar_ai_2023, hao_new_2023}.

Finally, we note that the scope of our quantitative analysis is limited to \textit{how} LLM research has been shifting, not \textit{why} these changes have occurred or how the field \textit{ought} to respond to them.
These latter questions are more well-suited to complementary qualitative methods, like interviews and surveys, which have also been a focus of recent work \citep{gururaja_build_2023, michael_what_2022}. 
In Section \ref{sec:discussion}, we pose four questions that future work can address by seeking out community perspectives. 
These follow-up discussions can help realize the value of the bibliometric findings we describe.

%% file: 99_appendix.tex
\setcounter{figure}{0}
\makeatletter 
\renewcommand{\thefigure}{S\@arabic\c@figure}
\makeatother

\setcounter{table}{0}
\makeatletter 
\renewcommand{\thetable}{S\@arabic\c@table}
\makeatother

\section{Appendix: Supplementary Methods}
\label{sec:supplementary_methods}

\begin{table*}[!htb]
    \centering
    \small
    \caption{A summary of data fields used in our analyses. Further details in \S \ref{sec:methods} and Appendix \ref{sec:supplementary_methods}.}
    \begin{tabular}{@{}p{3cm}p{3cm}p{8.8cm}@{}}
        \toprule
        & Field Name & Field Description \\
        \midrule
        \multirow{1}{*}{About LLMs? (\S \ref{sec:llm_related_definition})} & mentions keyword & 1 if paper title or abstract contains an LLM-related keyword from the list \{language model, foundation model, BERT, XLNet, GPT-2, GPT-3, GPT-4, GPT-Neo, GPT-J, ChatGPT, PaLM, LLaMA\}, 0 otherwise \\
        \midrule
        \multirow{2}{*}{Topics (\S \ref{sec:methods_topic_model})} & sub-arXiv & name of subarXiv paper belongs to (\textit{e.g.} Computation and Language) \\
        & topic & name of topic generated from topic model (\textit{e.g.} Code Generation) \\
        \midrule
        \multirow{2}{*}{Affiliations (\S \ref{sec:methods_industry_academic})} & academic & 1 if paper has $\ge 1$ academic affil.~with $\ge 10$ LLM papers, 0 otherwise \\
        & industry & 1 if paper has $\ge 1$ industry affil.~with $\ge 10$ LLM papers, 0 otherwise \\
        \midrule
        \multirow{1}{*}{Citations (\S \ref{sec:methods_citation_counts})}
        & citation percentile & citation percentile among papers published in same 3-month interval\\
        \bottomrule
    \end{tabular}
    \label{tab:merged_rows}
\end{table*}

\subsection{Data}\label{sec:methods_data}
We downloaded the latest release of arXiv metadata from Kaggle \cite{kaggle_leveraging_2020} in September, including all papers posted to the arXiv through 7 September 2023.\footnote{The arXiv dataset has a CC0 license, allowing for all use. Our annotated version of the dataset and our analysis code uses an MIT license, and is intended for further bibliometric research.} 
The metadata includes arXiv ID, author list, title, abstract, submission date, and arXiv subject categories. 
We subsetted the data to include only papers that listed in at least one CS or Stat sub-arXiv, and we further subset the data to only include papers since the start of 2018, resulting in \totalPapers papers.
(We chose 2018 to roughly align with the growing use of representations from pretrained language models, like BERT and ELMo \cite{devlin_bert_2019, peters_deep_2018}.) We downloaded PDF full-texts for all these papers, available from a GCP bucket jointly hosted by Kaggle and arXiv \cite{kaggle_leveraging_2020}. 
We then apply PDF-to-text conversion to produce plaintext files for each paper using the \texttt{pdftotext} Python tool. 
To study influential papers, we also pulled citation data using the Semantic Scholar API \cite{kinney_semantic_2023} for the \totalBroadLLMPapers LLM papers described below. Our repository includes documented code snippets for each of these steps, so future timepoints of arXiv data will also be easy to annotate. Key data fields which we use for analysis are summarized in Table \ref{tab:merged_rows}, with further explanation of each field below.

\subsection{Identifying LLM-related papers}\label{sec:llm_related_definition}
Consistent with past ML survey papers \cite{fan_bibliometric_2023, peng_mitigating_2021, blodgett_language_2020, field_survey_2021}, we composed an analysis subset of LLM papers by searching for an interpretable set of keywords.
Since we are interested in characterizing temporal trends, we intentionally chose a broad set of terms; in particular, we wanted to capture relevant papers \textit{before} modern instruction-tuned, chat-style LLMs, to better answer research questions about changes over time.
As such, we include terms like ``language model'' and ``BERT,'' which have been in use for longer than ``large language model.'' 
The complete keyword list consists of \{language model, foundation model, BERT, XLNet, GPT-2, GPT-3, GPT-4, GPT-Neo, GPT-J, ChatGPT, PaLM, LLaMA\},\footnote{We ignore case for ``language model'' and ``foundation model,'' and enforce case for the other keywords.} and \totalBroadLLMPapers papers since 2018 contain at least one of them in their title or abstract. 

We decided to include ``foundation models'' despite the fact that they can also refer to vision models; upon manually inspecting 50 out of the 521 papers mentioning this term, we found that the majority -- 38 out of 50 -- involved language at least to some extent (\textit{e.g.,} large vision-language models, LVLMs), so we included this term to capture the growing interest in multimodal approaches.
Besides ``language model'' and ``foundation model,'' we chose the specific model keywords by referencing Wikipedia's page for LLMs as of July 2023 \citep{wikipedia_large_2023}.
We removed the long tail of models for which there are $< 10$ hits (e.g., Chinchilla, LaMDA, Galactica) or many false positives (e.g., OPT, Claude, BLOOM -- for which there are many hits entirely unrelated to NLP). 
We inspected 50 paper abstracts which mention one of these less-common architectures and found that all of them use at least one other keyword on our list, so we expect minimal reduction to recall as a result of removing these long-tail keywords.
We refer to this entire set of \totalBroadLLMPapers papers as ``LLM-related papers'' and sometimes abbreviate to ``LLM papers,'' though it's important to note that not every paper in this set may adhere to the most recent (and evolving) conception of a ``large language model.'' 

To validate our keyword-based retrieval strategy, one study author manually annotated a random sample of 169 papers (1\% of the corpus) as either not relevant, somewhat relevant, or definitely relevant. 
``Not relevant'' papers were papers that entered the corpus spuriously, \textit{i.e.,} one of the keywords matched the abstract in a different word sense (like ``foundation modeling'' in seismology); ``somewhat relevant'' papers used one of the keywords correctly, but was mentioned as an example for a tangentially-related research topic (like how a new efficiency method can be useful for language models); ``definitely relevant'' papers used a language model or directly addressed them. Overall 153 (91\%) papers were definitely relevant, 15 papers were somewhat relevant, and 1 paper was not relevant. 

\subsection{Topic modeling}\label{sec:methods_topic_model} Several of our analyses rely on paper topic annotations, for example to identify sub-areas that are receiving increased research attention. 
We identifed paper topics both for the full set of \totalPapers CS/Stat arXiv papers, and the subsample of \totalBroadLLMPapers LLM-related papers.
We applied a recent topic modeling approach, using semantic text embeddings followed by dimensionality reduction and clustering. 
More specifically, we adopted the following workflow (closely resembling that of \cite{zhang_neural_2022}): 
(1) embed paper abstracts in a 768-dimensional space using the open source INSTRUCTOR-XL model \cite{su_one_2023};
(2) apply PCA to reduce dimensionality to $n=200$ components, which explain $\sim$90\% of the embedding variance; 
(3) apply UMAP \cite{mcinnes_umap_2020} to further transform the data into 2D-space while preserving local structure; 
(4) cluster the papers in 2D space using $k$-means or agglomerative clustering (discussed below); 
(5) assign an informative topic name to each cluster. We map clusters to topics one-to-one, as is standard \cite{zhang_neural_2022, grootendorst_bertopic_2022}, and we refer to them interchangeably. Recent work \cite{zhang_neural_2022, grootendorst_bertopic_2022, sia_tired_2020, thompson_topic_2020} shows that this embedding-based approach presents a more accurate and efficient alternative to other methods, such as LDA.

Clustering (step 4) varied slightly for the full set and the LLM subset: for the full set of \totalPapers CS/Stat papers we used $k$-means with 100 clusters, and for the subset of \totalBroadLLMPapers LLM papers we used Ward agglomerative clustering with 40 clusters.
While Ward and $k$-means yielded highly similar results (adjusted Rand index $> 0.6$), we preferred Ward for its improved adaptability to uneven cluster sizes; $k$-means, however, was the only method that scaled and yielded plausible results for the full paper set.
Choosing the cluster count $k$ required manual tuning to ensure that clusters were neither too broad nor overly redundant. 
For example, with $k=30$ clusters in the broad LLM subset, papers about language model stereotyping were in the same cluster as papers about predicting hate speech, while with $k=50$ there were multiple thematically similar clusters about low-resource languages, so we settled on $k=40$.

The final step of annotating clusters has previously been done by constructing a TF-IDF matrix and identifying the enriched terms that distinguish the cluster from others \cite{zhang_neural_2022, grootendorst_bertopic_2022}. 
In this approach, the researcher is left to manually synthesize the over-represented terms into succinct topic titles.
Though intuitive, the process of converting papers to terms and then terms to topic names adds an unnecessary step; instead, we prompt an LLM (\texttt{gpt-4} through the OpenAI API as of 15 Sep 2023) to use a sample of the cluster's paper titles and abstracts and, from those, directly assign a succinct cluster name. 
In many cases, these names were either too long, not specific enough, or overly specific.
We performed a manual pass by looking at samples of 25 papers per cluster to ensure that (a) the papers are thematically coherent and (b) the topic title is suitable (and we edited the titles for clarity/brevity, as necessary).

\subsection{Identifying industry and academic affiliations}\label{sec:methods_industry_academic}

Since author affiliations are rarely available in arXiv metadata, we extract affiliations by searching for regular expressions in paper full-texts, a common approach for metadata extraction in bibliometrics \cite{tkaczyk_new_2017, nasar_information_2018}. 
Specifically, many papers list author emails in the full-text, and we search for them with high precision by designing regexs to match the ``\texttt{@}'' symbol and appropriate surrounding text\footnote{We use two regexs, to match two possible formats: \texttt{author@domain.xyz} and \texttt{\{author1,author2,$\cdots$\}@domain.xyz.}
} in the paper's first 100 lines.
We conduct a manual audit of 100 papers to verify that all the extracted strings are author emails, and that the papers \textit{without} any extracted emails indeed do not obviously list emails in the manuscript.
Overall, 86\% of the LLM paper subset has at least one extracted email, which drops to 83\% after removing uninformative domains such as \texttt{gmail.com}.
Based on a manual audit of 50 papers, we did not observe that papers missing emails over-represent any particular type of affiliation.
However, we did find that most of the papers without emails listed affiliations in plain text, suggesting that more sophisticated metadata extraction methods may be able to extract a higher proportion of affiliations. 
We tried using one such tool, GROBID \citep{lopez_grobid_2009}, but it did not work as well as emails, so we leave this task to future work.

Using the list of emails associated with each paper, we labeled each paper depending on whether it has (1) an academic affiliation and (2) an industry affiliation. (Some papers may have both academic and industry affiliations and others may have none.) To perform this annotation, we extracted the domain name (e.g., `cornell.edu') from each email. We then combined domain names that correspond to the same institution using a semi-automated approach. Furthermore, for all domains $d$ with at least 10 papers in our entire arXiv dataset, we mapped all domains $s$ that are subdomains of $d$ to $d$ (e.g., $s=$`cs.cornell.edu'$\mapsto$ $d=$`cornell.edu'). We also manually identified groups of domains with at least 10 LLM papers (e.g., `fb.com' and `meta.com'). 
Finally, among the 329 remaining domains with at least 10 LLM papers, we manually coded them as either academic domains or industry domains.
Two study authors independently performed this task and resolved any disagreements through discussion.

\subsection{Citation counts}\label{sec:methods_citation_counts}

For each of the LLM papers, we pull its citation count from the Semantic Scholar API \cite{kinney_semantic_2023}, reflecting the number of times another paper has cited it (note that this includes \textit{all} references, not just those by other LLM papers). 
Because papers in 2018, for example, have had more time to accrue citations than papers from 2022, we avoid using raw citation counts.
Instead, we compute each paper's 3-month \emph{citation percentile}: that is, the percentile of its citation count comparing only to other papers released during the same 3-month interval, so \textit{e.g.} a paper from 2020-02-15 would be compared only to papers from 2020-01-01 through 2020-03-31. 
Percentile-based metrics for citations are commonly used in bibliometric analysis~\cite{konkiel_how_2021}.
Each paper is mapped to one of 22 intervals: [2018-01-01, 2018-04-01], $\dots$, [2023-04-01, 2023-07-01].
We exclude papers from the most recent 3-month interval, leaving 15,147 papers in this analysis.
Note that Semantic Scholar caps the number of tracked citations at 10,000 per paper; only 3 papers in our subset are above this threshold, so it does not meaningfully affect our results.

\section{Studying first-time authors using arXiv publication history}

\subsection{Identifying new \& experienced authors}\label{sec:methods_author_background}
New authors are defined as authors who have not previously co-authored any NLP paper, i.e., a paper in cs.CL. 
We chose this definition because the proportion of new authors is then comparable year-over-year. 
In particular, defining new authors according to whether they have previously written any other LLM paper prevents such comparisons, because LLM research (according to our definition) only begins in 2018. 
An author in 2021, for example, would have had less time to publish an LLM paper than an author in 2023.
Despite this inability to compare year-over-year, we also computed the percentage of unique first authors in 2023 that have not previously published an LLM paper, which is 66\%. (Understandably, this percentage is higher than the percent who have never previously written an NLP paper: the field of LLM research is much smaller than NLP as a whole, so not as many authors have already written an LLM paper.)

Authors are matched using their names as entered on arXiv. 
It is possible that this mis-estimates the fraction of first-time authors, due to imperfect linkage across papers.
Authors could be incorrectly marked as first-time authors if their name on arXiv changed, or they could be incorrectly marked as experienced authors if multiple authors have the same name.
Because we were most concerned with the relative statistics (i.e., whether new author percentage was notably higher in 2023 than 2022), we did not take any steps to mitigate these issues.
However, it is important to note that these variables may contain some errors.

\subsection{Publication histories of new \& experienced authors}
\label{sec:pub_history}

Among the 2023 LLM papers, there are 2,746 unique first authors who are new (\textit{i.e.} no prior NLP paper), and 2,796 unique first authors who are experienced.
Out of the new authors, 1,570 (57\%) of them had at least one paper before their first LLM paper, while the rest of them had no prior papers (suggesting that they are junior researchers). 
By definition, all experienced authors had at least one paper in cs.CL (and potentially other papers) before their first LLM paper.

To study new authors' disciplinary backgrounds, we study the sum total of their arXiv histories prior to 2023. 
Collectively, they have written 9,476 papers on arXiv; we compare the sub-arXiv distribution over these papers to the 51,107 papers written by experienced authors.
Table \ref{tab:past-subarxivs} lists the top-10 sub-arXivs, sorted by new author paper count.
Over half (51\%) of the papers previously written by new authors are in Computer Vision or Machine Learning, which are common categories overall (accounting for 36\% of experienced-author papers).
But a long tail of less common sub-arXivs further distinguishes new and experienced authors: new authors have published more in Software Engineering, Robotics, Security, and Social Networks, which are the same sub-arXivs growing in overall LLM proportion in 2023 (\S\ref{sec:results_part_b_what_topics}).
We conclude that, in 2023, many researchers with prior experience in other fields are entering LLM research.

\newpage

\begin{figure*}[!ht]
    \centering
    \includegraphics{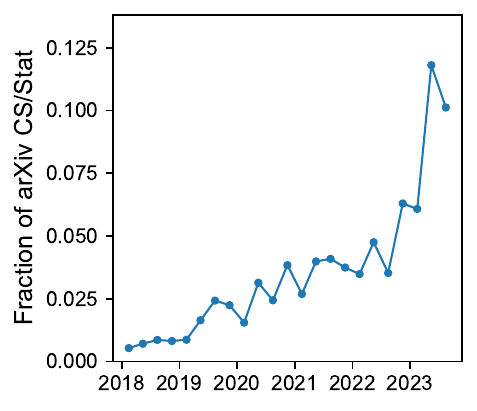}
    \caption{The overall incidence of LLM-related papers has increased substantially in the last few years, up to 12\% of all arXiv CS/Stat submissions since the second quarter of 2023. 
    Papers are identified as LLM-related if their title or abstract contains one of the keywords listed in \S\ref{sec:llm_related_definition}.
    }
\label{fig:supp_percent_llm_arxiv}
\end{figure*}

\begin{figure*}[!ht]
    \centering
    \includegraphics[width=\suppEnrichmentFigWidth\textwidth]{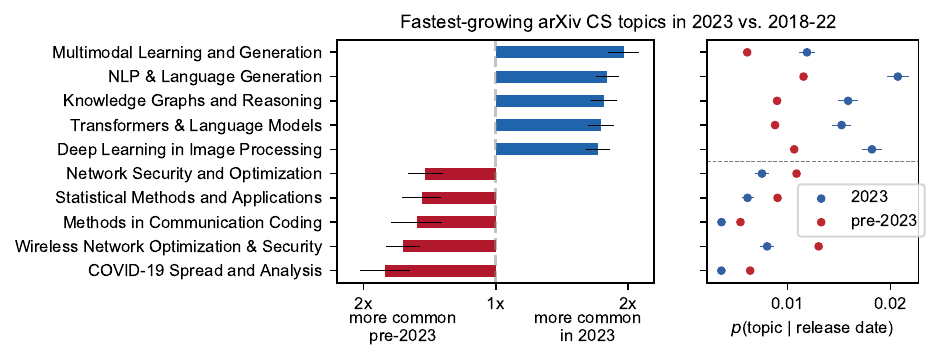}
    \caption{The fastest-growing and shrinking topics out of all 100 annotated topics on the CS/Stat arXivs. 
    All of the top four topics involve multimodal models, NLP, and LLMs. 
    These results suggest that LLMs have grown quickly not only in terms of absolute statistics, but also relative to all other growing topics within CS research.
    }
\label{fig:supp_top_arxiv_cs_topics}
\end{figure*}

\begin{table*}[htbp]
\small
\centering
\caption{The 50 most disproportionately-used keywords (including unigrams \& bigrams) in arXiv CS/Stat paper abstracts in 2023 vs.~2018--2022. Most of the keywords relate to LLMs, though some terms also relate to Segment Anything (SAM) and multimodal text-image models (\textit{e.g.}, CLIP). Keywords that never appear prior to 2023 are excluded.}
\begin{tabular}{lr}
\toprule
                         Keyword &  $\frac{p(\text{keyword} | \text{post-2023})}{p(\text{keyword} | \text{pre-2023})}$ \\
\midrule
                   \texttt{2023} &                                              189.4 \\
                  \texttt{llama} &                                              188.6 \\
                \texttt{chatgpt} &                                              179.8 \\
           \texttt{like chatgpt} &                                              132.6 \\
              \texttt{llm based} &                                               80.8 \\
          \texttt{segment model} &                                               73.1 \\
            \texttt{models llms} &                                               57.4 \\
                   \texttt{llms} &                                               54.4 \\
                    \texttt{llm} &                                               46.4 \\
       \texttt{stable diffusion} &                                               44.1 \\
              \texttt{model llm} &                                               43.0 \\
        \texttt{image diffusion} &                                               41.2 \\
                    \texttt{cot} &                                               41.1 \\
      \texttt{llms demonstrated} &                                               35.6 \\
             \texttt{llms shown} &                                               35.6 \\
          \texttt{generative ai} &                                               35.4 \\
          \texttt{chain thought} &                                               29.5 \\
      \texttt{foundation models} &                                               23.9 \\
         \texttt{large language} &                                               21.5 \\
           \texttt{text prompts} &                                               16.0 \\
       \texttt{diffusion models} &                                               12.4 \\
       \texttt{foundation model} &                                               12.2 \\
       \texttt{latent diffusion} &                                               12.1 \\
        \texttt{diffusion model} &                                               11.3 \\
           \texttt{tuned models} &                                               11.1 \\
     \texttt{prompt engineering} &                                               11.0 \\
            \texttt{underscores} &                                                9.9 \\
      \texttt{valuable insights} &                                                9.8 \\
             \texttt{underscore} &                                                9.8 \\
            \texttt{models clip} &                                                9.6 \\
    \texttt{denoising diffusion} &                                                9.6 \\
                \texttt{prompts} &                                                8.1 \\
  \texttt{instruction following} &                                                8.1 \\
            \texttt{text guided} &                                                7.9 \\
           \texttt{ai generated} &                                                7.9 \\
            \texttt{text prompt} &                                                7.7 \\
       \texttt{context learning} &                                                7.6 \\
              \texttt{prompting} &                                                7.6 \\
        \texttt{open vocabulary} &                                                7.4 \\
                    \texttt{sam} &                                                7.2 \\
        \texttt{diffusion based} &                                                7.0 \\
                    \texttt{gpt} &                                                7.0 \\
                 \texttt{prompt} &                                                6.8 \\
                  \texttt{nerfs} &                                                6.6 \\
\texttt{demonstrated remarkable} &                                                6.6 \\
             \texttt{models gpt} &                                                6.6 \\
         \texttt{hallucinations} &                                                6.5 \\
\bottomrule
\end{tabular}
\label{tab:keyword_enrichment}
\end{table*}

\include{tables/supp_table_fastest_growing_llm_2023}

\begin{figure*}[!htb]
    \begin{center}  \includegraphics[width=\suppEnrichmentFigWidth\textwidth]{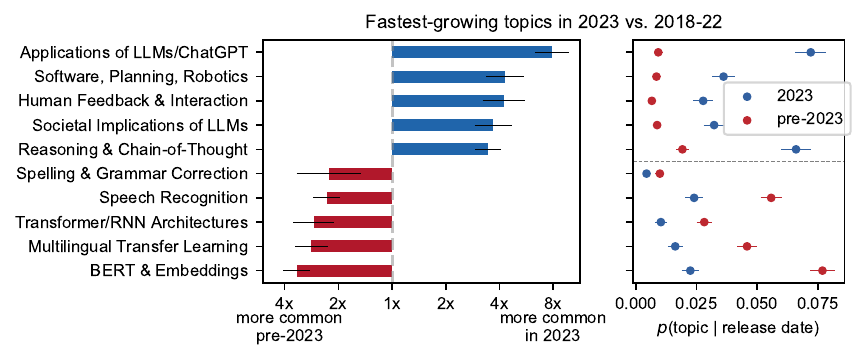}
    \end{center}
    \caption{The fastest-growing LLM research topics in 2023. LLM papers increasingly focus on applications, task evaluations, and prompting, while papers on BERT and model architecture are becoming less common. Left: Enrichment ratios given by $\frac{p(\text{topic} | \text{2023})}{ p(\text{topic} | \text{pre-2023})}$. Blue topics are more common in 2023, red topics are more common pre-2023. Right: Topic frequencies pre-2023 and since 2023. 
    }
    \label{fig:fast-growing-topics}
\end{figure*}

\begin{figure*}
    \begin{center}  \includegraphics[width=\textwidth]{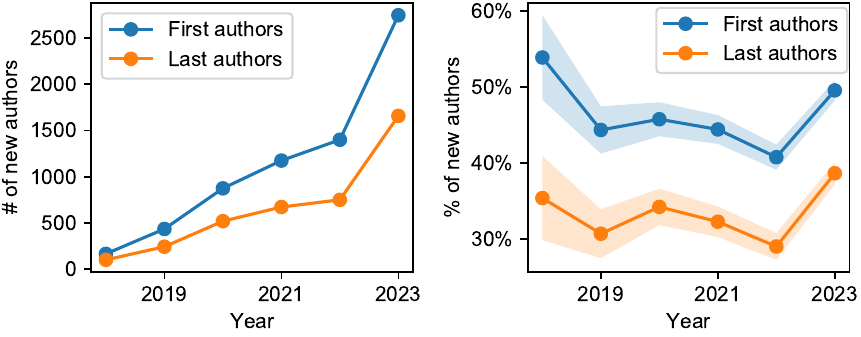}
    \end{center}
    \caption{\textbf{Counts (left) and percentages (right) of LLM papers with new first and last authors, 2018-2023.} 
    Here, ``new authors'' are defined as those who never previously co-authored an NLP paper on arXiv prior to their LLM paper. 
    Left: The count of new authors rose at a roughly linear rate as the field grew, but then jumped significantly in 2023. (Also note that the count in 2023 is skewed lower, because our corpus only includes papers until Sep.~7, 2023.)
    Right: The percentage of new authors was high in 2018, decreased and stagnated in 2019-22, and has once again risen significantly in 2023. These trends reflects that the recent ChatGPT era of LLMs -- like the era five years ago after the release of BERT and ELMo -- has brought in a large number of new authors. 
    }
    \label{fig:lineplot-new-authors}
\end{figure*}

\include{tables/supp_table_2023_newauth_llm_subarxivs}

\begin{figure*}
    \begin{center}  \includegraphics[width=\suppEnrichmentFigWidth\textwidth]{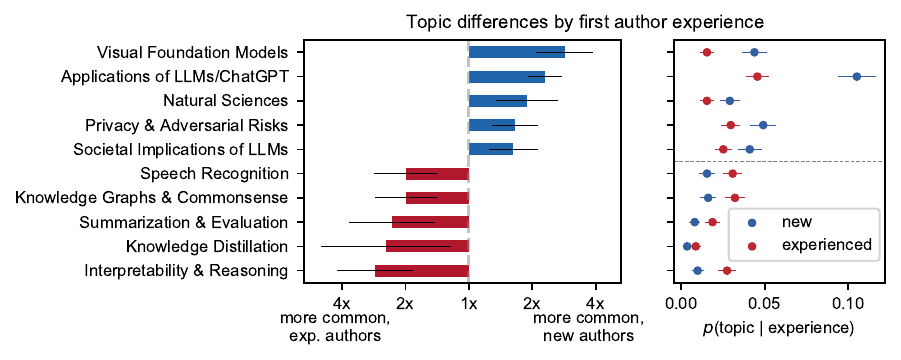}
    \end{center}
    \caption{Topics of LLM papers written in 2023 vary with first author experience. Topics in blue are more common for papers whose first author is new (has never written an NLP paper prior to their first LLM paper), while topics in red are more common for papers with experienced authors.
    }
    \label{appendix:topics_experience_first_author}
\end{figure*}

\begin{figure*}[!ht]
    \begin{center}        \includegraphics[width=\textwidth]{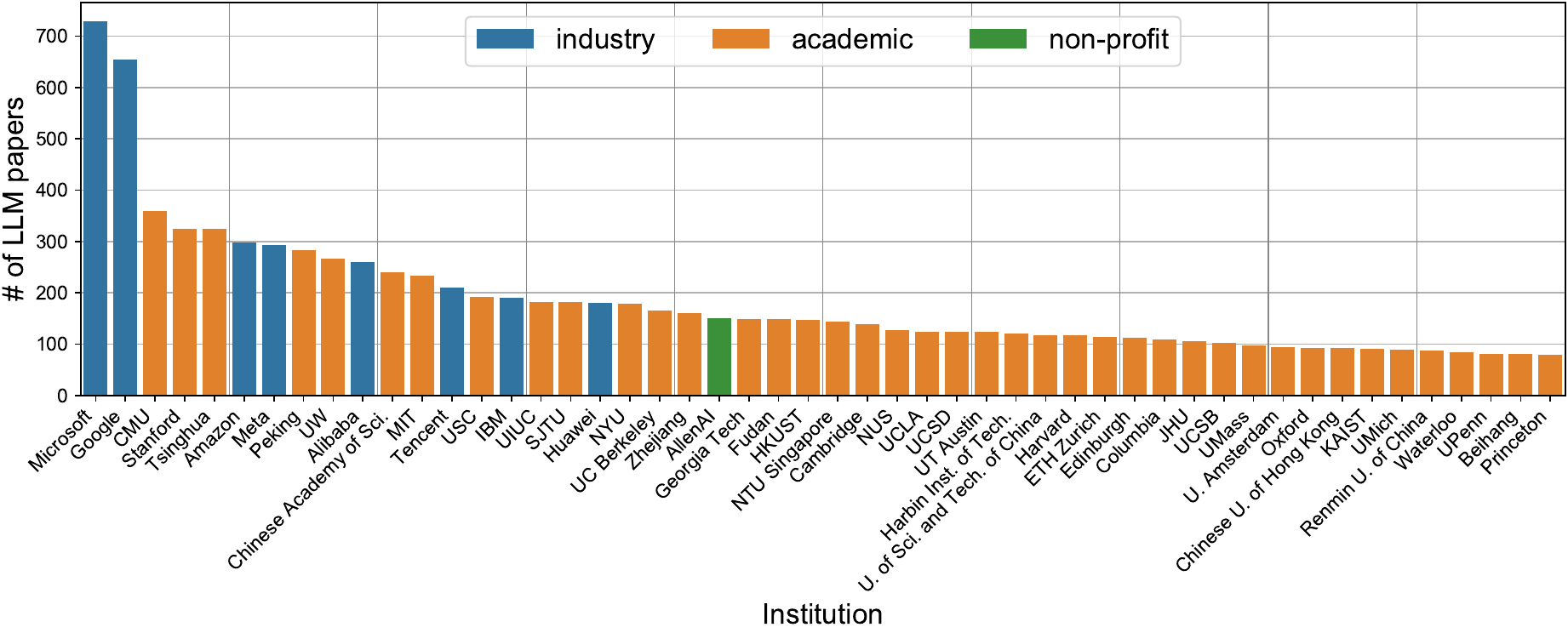}
    \caption{The 50 institutions with the most LLM papers. Most are academic, but there are several large industry players, with Microsoft and Google producing by far the most papers.}
    \label{fig:top_institutions}
    \end{center}
\end{figure*}

\begin{figure*}
    \centering
    \includegraphics[width=\enrichmentFigWidth\textwidth]{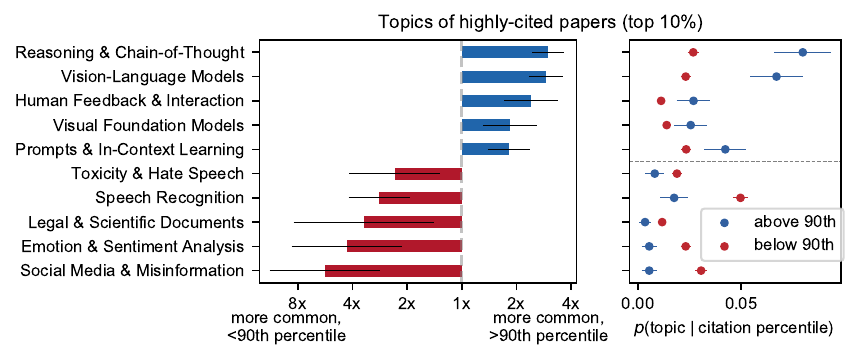}
    \caption{\textbf{Highly-cited papers often report on new models and improved performance.} Interestingly, three of the topics unlikely to be highly-cited (Toxicity \& Hate Speech; Legal \& Scientific Documents; Social Media \& Misinformation) are also three of the topics which are more common in academic papers. Bar plot displays the ratio $\frac{p(\textrm{topic}|\textrm{above 90th percentile})}{p(\textrm{topic}|\textrm{below 90th percentile})}$. Dot plot displays the frequency of the topic in the highly-cited (blue) and not highly-cited (red) groups.}
    \label{fig:highly_cited_topics}
\end{figure*}

%% file: tables/supp_table_fastest_growing_llm_2023.tex
\begin{table*}
\small
\centering

\caption{\textbf{Topics in the LLM literature that have grown and shrunk the most in 2023, compared to 2018-2022.} $p$-values computed with a $\chi^2$ test. The 40 topics were assigned and labeled using our neural topic modeling approach described in the methods. Topics are sorted by $\frac{p(\text{topic} \mid \text{since-2023})}{p(\text{topic} \mid \text{pre-2023})}$.}

\label{appendix:supp-enriched-topics-2023}

\begin{tabular}{lcccccc}
\toprule
                                 Topic &   $N$ & $\frac{p(\text{topic} \mid \text{since-2023})}{p(\text{topic} \mid \text{pre-2023})}$ & $p(\text{topic} \mid \text{since-2023})$ & $p(\text{topic} \mid \text{pre-2023})$ & $p$-value \\
\midrule
        Applications of LLMs/ChatGPT & 560 &                                               7.84 &                                    0.072 &                                  0.009 &  3.7e-109 \\
        Software, Planning, Robotics & 321 &                                               4.28 &                                    0.036 &                                  0.008 &   2.1e-37 \\
        Human Feedback \& Interaction & 247 &                                               4.24 &                                    0.028 &                                  0.007 &   1.2e-28 \\
       Societal Implications of LLMs & 299 &                                               3.69 &                                    0.032 &                                  0.009 &   3.0e-29 \\
        Reasoning \& Chain-of-Thought & 626 &                                               3.45 &                                    0.066 &                                  0.019 &   2.7e-55 \\
            Visual Foundation Models & 270 &                                               3.28 &                                    0.028 &                                  0.009 &   1.4e-22 \\
                Finance Applications & 110 &                                               1.83 &                                    0.009 &                                  0.005 &   1.8e-03 \\
              Vision-Language Models & 504 &                                               1.72 &                                    0.040 &                                  0.023 &   4.9e-10 \\
      Applications \& Benchmark Evals & 336 &                                               1.64 &                                    0.026 &                                  0.016 &   4.6e-06 \\
     Fine-Tuning \& Domain Adaptation & 364 &                                               1.62 &                                    0.028 &                                  0.017 &   3.1e-06 \\
           Video \& Multimodal Models & 437 &                                               1.56 &                                    0.033 &                                  0.021 &   2.5e-06 \\
                    Natural Sciences & 296 &                                               1.43 &                                    0.021 &                                  0.015 &   2.1e-03 \\
         Privacy \& Adversarial Risks & 528 &                                               1.43 &                                    0.038 &                                  0.027 &   3.2e-05 \\
                     Code Generation & 408 &                                               1.42 &                                    0.029 &                                  0.021 &   4.3e-04 \\
              Audio \& Music Modeling & 119 &                                               1.41 &                                    0.009 &                                  0.006 &   7.3e-02 \\
       Prompts \& In-Context Learning & 418 &                                               1.22 &                                    0.028 &                                  0.023 &   4.9e-02 \\
                  NLP for Healthcare & 697 &                                               1.20 &                                    0.046 &                                  0.038 &   1.4e-02 \\
        Interpretability \& Reasoning & 315 &                                               1.09 &                                    0.020 &                                  0.018 &   4.6e-01 \\
                      Biases \& Harms & 448 &                                               1.01 &                                    0.027 &                                  0.026 &   9.3e-01 \\
        Legal \& Scientific Documents & 179 &                                               0.91 &                                    0.010 &                                  0.011 &   6.1e-01 \\
          Entity Extraction \& RecSys & 435 &                                               0.87 &                                    0.023 &                                  0.027 &   1.9e-01 \\
          Translation \& Low-Resource & 433 &                                               0.81 &                                    0.022 &                                  0.027 &   4.1e-02 \\
      Knowledge Graphs \& Commonsense & 492 &                                               0.80 &                                    0.025 &                                  0.031 &   2.0e-02 \\
            Efficiency \& Performance & 987 &                                               0.73 &                                    0.047 &                                  0.065 &   4.1e-06 \\
          Summarization \& Evaluation & 303 &                                               0.69 &                                    0.014 &                                  0.020 &   3.8e-03 \\
        Dialogue \& Conversational AI & 516 &                                               0.68 &                                    0.023 &                                  0.035 &   5.4e-05 \\
               Datasets \& Benchmarks & 238 &                                               0.66 &                                    0.011 &                                  0.016 &   3.7e-03 \\
                  Search \& Retrieval & 341 &                                               0.63 &                                    0.015 &                                  0.023 &   1.5e-04 \\
      Question Answering \& Retrieval & 611 &                                               0.61 &                                    0.026 &                                  0.042 &   3.7e-08 \\
       Social Media \& Misinformation & 458 &                                               0.56 &                                    0.018 &                                  0.032 &   2.8e-08 \\
                     Text Generation & 424 &                                               0.54 &                                    0.016 &                                  0.030 &   2.5e-08 \\
        Emotion \& Sentiment Analysis & 347 &                                               0.53 &                                    0.013 &                                  0.025 &   2.9e-07 \\
              Knowledge Distillation & 169 &                                               0.53 &                                    0.006 &                                  0.012 &   3.4e-04 \\
Pretrained LMs \& Text Classification & 674 &                                               0.52 &                                    0.025 &                                  0.049 &   3.3e-14 \\
              Toxicity \& Hate Speech & 290 &                                               0.51 &                                    0.011 &                                  0.021 &   8.7e-07 \\
       Spelling \& Grammar Correction & 132 &                                               0.44 &                                    0.004 &                                  0.010 &   1.1e-04 \\
                  Speech Recognition & 742 &                                               0.43 &                                    0.024 &                                  0.056 &   1.4e-22 \\
       Transformer/RNN Architectures & 363 &                                               0.36 &                                    0.010 &                                  0.028 &   8.7e-15 \\
      Multilingual Transfer Learning & 587 &                                               0.35 &                                    0.016 &                                  0.046 &   2.1e-24 \\
                   BERT \& Embeddings & 955 &                                               0.29 &                                    0.022 &                                  0.077 &   3.2e-50 \\
\bottomrule
\end{tabular}
\end{table*}

%% file: tables/supp_table_2023_newauth_llm_subarxivs.tex
\begin{table*}
\small
\centering

\caption{\textbf{Sub-arXiv publishing histories of the new LLM authors in 2023.} 
The table compares the publication histories of the 9,476 papers written by new LLM authors to the publication histories of the 51,107 experienced LLM authors, prior to 2023. 
New authors have historically published more in both general ML sub-arXivs and other specific sub-fields. 
The table only counts the primary sub-arXiv designation. Bold values denote fractions which are significantly larger in one group than the other, \textit{i.e.}, $p < 0.005$ under a $\chi^2$ test. 
}

\label{tab:past-subarxivs}

\begin{tabular}{lccc}
\toprule
                sub-arXiv &         \% new & \% experienced \\
\midrule
    Computation and Language & 0.0 & \textbf{23.8} \\
\midrule
          Computer Vision & \textbf{30.4} &          20.7 \\
         Machine Learning & \textbf{20.9} &          15.4 \\
     Software Engineering &  \textbf{3.8} &           1.5 \\
       Information Theory &           3.7 &           3.8 \\
                 Robotics &  \textbf{3.5} &           2.1 \\
Cryptography and Security &  \textbf{3.1} &           2.2 \\
  Artificial Intelligence &           2.6 &           2.3 \\
Social and Info. Networks &  \textbf{2.3} &           1.4 \\
    Information Retrieval &           2.1 &  \textbf{3.1} \\
                Stat - ML &           1.8 &           1.5 \\

\bottomrule
\end{tabular}
\end{table*}

\begin{table*}
\small
\centering

\caption{\textbf{In 2023, new \& experienced authors are writing LLM papers on different topics.} Sub-arXiv distributions for the LLM-related papers written by new first authors (those without a prior NLP paper) vs. experienced first authors.
New authors are writing less in cs.CL and more in other sub-arXivs.
The top 10 sub-arXivs by new author count are listed here; $p$-values computed with a $\chi^2$ test.}

\label{supp:tab_2023_newauth_llm_subarxiv}

\begin{tabular}{lccccc}
\toprule
                 sub-arXiv &  count, new &  count, experienced & \% new & \% experienced & $p$-value \\
\midrule
  Computation and Language &        1102 &                2364 &  38.6 &          66.1 &   3.9e-34 \\
           Computer Vision &         435 &                 303 &  15.2 &           8.5 &   7.7e-14 \\
          Machine Learning &         299 &                 213 &  10.5 &           6.0 &   1.2e-09 \\
   Artificial Intelligence &         174 &                 153 &   6.1 &           4.3 &   2.1e-03 \\
      Software Engineering &         163 &                 101 &   5.7 &           2.8 &   4.0e-08 \\
     Computers and Society &         108 &                  34 &   3.8 &           1.0 &   1.2e-13 \\
Human-Computer Interaction &         102 &                  44 &   3.6 &           1.2 &   1.6e-09 \\
     Information Retrieval &          79 &                 121 &   2.8 &           3.4 &   1.9e-01 \\
 Cryptography and Security &          77 &                  35 &   2.7 &           1.0 &   4.6e-07 \\
                  Robotics &          65 &                  37 &   2.3 &           1.0 &   1.5e-04 \\
\bottomrule
\end{tabular}
\end{table*}